\documentclass[%
 reprint,
 amsmath,amssymb,
 aps,
]{revtex4-2}

\usepackage{graphicx}
\usepackage{bm}
\usepackage{hyperref}
\usepackage{orcidlink}

\begin{document}
\preprint{APS/123-QED}

\title{Quantum Dissipative Continuous Time Crystals}
\author{Felix Russo\orcidlink{0009-0005-8975-2245}}
 \email{felix.russo@tuwien.ac.at}
\author{Thomas Pohl}%
 \email{thomas.pohl@itp.tuwien.ac.at}
\affiliation{%
Institute for Theoretical Physics, Vienna University of Technology (TU Wien), Vienna, Austria
}%
\date{\today}

\begin{abstract}
Continuous time crystals, i.e., nonequilibrium phases with a spontaneously broken continuous time-translational symmetry, have been studied and recently observed in the long-time dynamics of open quantum systems. Here, we investigate a lattice of interacting three-level particles and find two distinct time-crystal phases that cannot be described within mean-field theory. Remarkably, one of them emerges only in the presence of correlations, upon accounting for beyond-mean-field effects. Our findings extend explorations of continuous time-translational symmetry breaking in dissipative systems beyond the classical phenomenology of periodic orbits in a low-dimensional nonlinear system. The proposed model applies directly to the laser-driven dynamics of interacting Rydberg states in neutral atom arrays and suggests that the predicted time-crystal phases are observable in such experiments. 
\end{abstract}

\maketitle
The concept of symmetry breaking is instrumental in understanding a broad range of fundamental processes in physics, chemistry, and biology \cite{strocchi2005symmetry}. Among the most ubiquitous examples is the emergence of ordered structures from an otherwise homogeneous system. In recent years, the question of whether translational symmetry could be broken in the fourth dimension has attracted much attention \cite{wilczek2012quantum,sacha2017time,hannaford2022decade,zaletel2023colloquium}. A quantum system that self-organizes in time, i.e., a time crystal, represents a new exotic state of matter. Early theoretical work focused on discrete time-translational symmetry breaking in periodically driven systems \cite{sacha2015modeling,else2016floquet,khemani2016phase}, which has subsequently been observed in different experimental platforms \cite{choi2017observation,zhang2017observation,kyprianidis2021observation,pal2018temporal,rovny2018observation,o2020signatures,randall2021many,kessler2021observation,greilich2024exploring}.
For time-independent systems, the possibility of continuous time-translational symmetry breaking was pointed out in \cite{iemini2018boundary} in the form of a so-called boundary time crystal, where a fraction of a many-body system can undergo spontaneous oscillations. The emergence of a continuous time crystal (CTC) phase in this setting is directly connected to the physics of dissipative quantum systems, which can exhibit limit-cycle solutions \cite{iemini2018boundary,chan2015limit,kessler2019emergent,buvca2019non} that were recently observed experimentally in nonlinear optical cavities \cite{kongkhambut2022observation,li2024time}, optical metamaterials \cite{Liu2023}, thermal Rydberg vapors 
\cite{wadenpfuhl2023emergence,wu2024dissipative}, and semiconductors \cite{Greilich2024}.

Despite the diversity of physical platforms and models, an overarching question has been whether time-translational symmetry breaking can be understood classically and to what extent quantum mechanics determines the properties of time-crystal phases \cite{iemini2018boundary,zhu2019dicke,buvca2019non,Pizzi2021,Ye2021,Pizzi2021b,solanki2025generation}. In current experiments and theoretical models, the emergence of CTC phases in dissipative quantum many-body systems is largely understood in terms of a mean-field picture. 
Hereby, the dynamics of single-particle observables are governed by a small set of nonlinear equations of motion that can exhibit limit-cycle attractors.
In fact, the emergence of persistent oscillations in open quantum systems can be associated with an approach to the classical limit \cite{dutta2025}, while quantum-mechanical extensions of mean-field models have been found to cause damping of limit cycles \cite{owen2018quantum,zhu2019dicke}. Hence, there is currently no experimental or theoretical evidence for continuous quantum time crystal (qCTC) phases in open quantum systems that do not pertain to the limit-cycle phenomenology of nonlinear mean-field equations. 

\begin{figure}[t!]
\includegraphics[width=0.95\columnwidth]{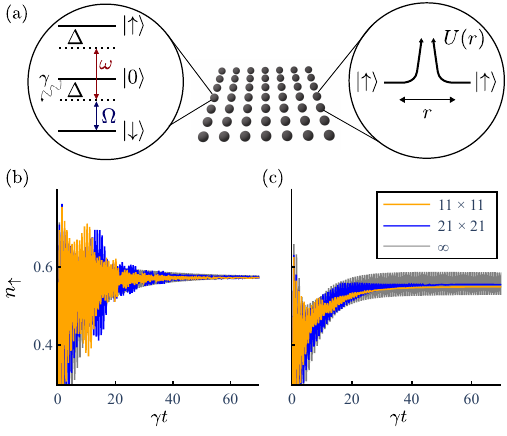}
\caption{\label{fig:setup} (a) The three states of spin$-1$ particles in a two-dimensional lattice are coupled by a resonant field with Rabi frequency $\omega$ and a field with Rabi frequency $\Omega$ that is detuned by $\Delta$ from the lower transition. Dissipation arises from spontaneous decay of the intermediate state with a rate $\gamma$. Interactions $U(r)$ cause state-dependent energy shifts for two particles at a distance $r$. The interplay of coherent coupling, dissipation, and interactions can lead to persistent oscillations. While mean-field theory predicts a stationary steady-state value of the $\lvert\uparrow\rangle$-state population $n_\uparrow$ in the thermodynamic limit for the chosen set of parameters in panel (b), persistent oscillations emerge upon accounting for beyond-mean-field effects as shown in panel (c). Such a quantum dissipative time-crystal phase, thus, breaks the continuous time translational symmetry of the system in the absence of a limit cycle in the classical mean-field dynamics.}
\end{figure}
\begin{figure*}
\includegraphics[width=\linewidth]{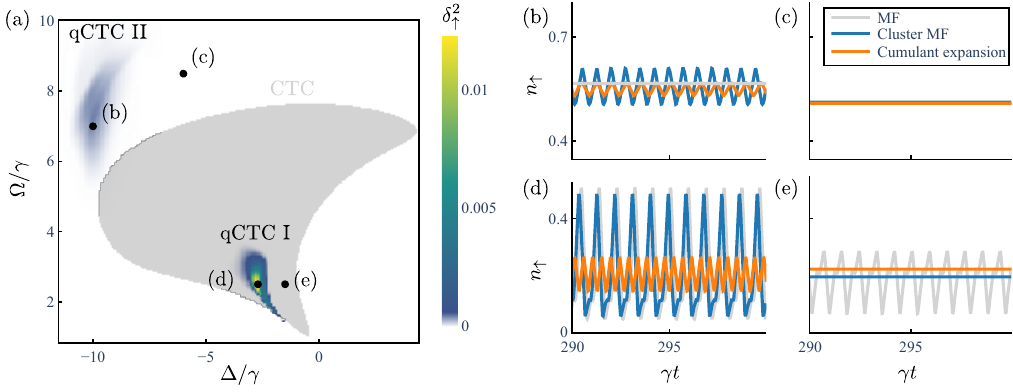}
\caption{\label{fig:pd} (a) Phase diagram in the thermodynamic limit for $\omega = 6 \gamma$ and $\chi = -18 \gamma$. The occurrence of mean-field CTCs is marked by the gray shaded area. The superimposed color map shows the asymptotic oscillation contrast $\delta^2_\uparrow$ (see Eq.(\ref{eq:osc})) from beyond-mean-field calculations, revealing two distinct oscillatory regions, which we term type-I and type-II qCTC. The qCTC-I phase resides within the mean-field CTC region, while the mean-field CTC is otherwise damped by beyond-mean-field effects, as illustrated in panels (d) and (e), respectively. In contrast, time-translational symmetry breaking in the qCTC-II phase occurs outside the classical CTC region and is driven by correlations, as shown in panels (b) and (c). The oscillation contrast in panel (a) is obtained from a $2^{\rm nd}$-order cumulant expansion, while panels (b)-(e) compare mean-field dynamics to the results of the cumulant expansion and cluster mean-field simulations described in the text.}
\end{figure*}
Here, we address this question by exploring the dissipative dynamics of a spin-1 lattice with finite-range interactions that do not afford an exact mean-field description. Numerical simulations reveal two distinct time crystalline phases with persistent oscillations of all observables in the thermodynamic limit. The first kind is found under conditions for which limit-cycle dynamics are predicted by mean-field theory but survive in the presence of significant quantum fluctuations. In addition, we identify a second type of qCTC that is not predicted by mean-field theory but instead is driven by correlations. This phase features a distinct scaling of quantum fluctuations with system size and emerges in the absence of long-range order, consistent with the phenomenology of oscillatory solutions in quenched isolated quantum systems \cite{medenjak2020isolated}. The found qCTC phases do not rely on symmetries of the underlying master equation and are robust to perturbations, both of which are essential aspects for potential experiments. Indeed, the studied model and considered parameters for qCTC should be realizable in current quantum simulation experiments with Rydberg atomic arrays. 

We consider a two-dimensional cubic lattice of three-level systems with quantum states $\lvert\downarrow\rangle$, $\lvert\uparrow\rangle$, and $\lvert0\rangle$, as illustrated in Figure~\ref{fig:setup}(a). The intermediate state $\lvert0\rangle$ is coupled to the upper $\lvert\uparrow\rangle$ and lower $\lvert\downarrow\rangle$ states with respective Rabi frequencies $\omega$ (resonant) and $\Omega$ ($\Delta$-detuned). The corresponding Hamiltonian for $N$ such three-level systems can be written as
\begin{align}
\label{eq:Hamiltonian}
\begin{split}
    \hat{H} 
        &= \Delta\sum_{i}  \hat{n}_\downarrow^{(i)} + \sum_i \left(\Omega \hat{\sigma}_{\downarrow0}^{(i)} +\omega \hat{\sigma}_{\uparrow0}^{(i)} + {\rm h.c.}\right) \\
        &\hphantom{=}+\sum_{i < j} U_{ij} \hat{n}_\uparrow^{(i)}\hat{n}_\uparrow^{(j)},
\end{split}
\end{align}
where $\hat{\sigma}_{\alpha\beta}^{(i)} = \lvert\alpha\rangle_i\langle\beta\rvert$ denote the transition operators for the $i^{\rm th}$ particle, and $\hat{n}_{\alpha}^{(i)} = \hat{\sigma}_{\alpha\alpha}^{(i)}$ corresponds to the respective excitation density at a given site. The second line in Eq.(\ref{eq:Hamiltonian}) describes density-density interactions between particles in the uppermost state $\lvert\uparrow\rangle$. We consider power-law potentials $U_{ij}=C_\kappa/r_{ij}^\kappa$, where $C_\kappa$ is the interaction strength and $r_{ij}=|{\bf r}_j-{\bf r}_i|$ the distance between two particles $i$ and $j$ at positions ${\bf r}_i$ and ${\bf r}_j$ in the lattice. We parameterize the interaction strength by the total interaction energy $\chi = \sum_{j \neq 0} U_{0j}$ of the particle at position ${\bf r}_0$ in the center of the lattice with all surrounding spins. 

This Hamiltonian applies directly to current experiments with neutral-atom arrays \cite{Labuhn2016b,bernien2017probing,scholl2021,browaeys2020many,kaufman2021quantum,srakaew2023subwavelength,weckesser2024realization}, where $\lvert\downarrow\rangle$ corresponds to an atomic ground state, $\lvert0\rangle$ to a low-lying intermediate state, and $\lvert\uparrow\rangle$ to a highly excited Rydberg state. Laser fields can be used to drive the $\lvert0\rangle\leftrightarrow\lvert\downarrow\rangle$ and $\lvert0\rangle\leftrightarrow\lvert\uparrow\rangle$ transitions with Rabi frequencies $\Omega$ and $\omega$, as described by Eq.(\ref{eq:Hamiltonian}). Considering resonant laser driving of the upper transition, the frequency detuning on the lower transition corresponds to the energy mismatch $\Delta$ in Eq.(\ref{eq:Hamiltonian}). Moreover, Rydberg atoms feature strong van der Waals interactions, $U_{ij}=C_6/r_{ij}^6$, \cite{gallagher1994rydberg,saffman2010quantum,weber2017calculation} while longer ranged dipole-dipole interactions, $U_{ij}=C_3/r_{ij}^3$, can be realized and controlled via microwave dressing of the Rydberg state \cite{tanasittikosol2011microwave,sevinccli2014microwave,kurdak2024enhancement}. While Rydberg states ($\lvert\uparrow\rangle$) typically feature very long lifetimes \cite{gallagher1994rydberg,saffman2010quantum,low2012experimental,morgado2021quantum}, the radiative decay of the intermediate $\lvert0\rangle$-state introduces dissipation as described by the Lindbladian
\begin{align}
\label{eq:Lindbladian}
\begin{split}
    \hat{\mathcal{L}}(\hat{\rho}) = \gamma \sum_{i} \left( \hat{\sigma}_{\downarrow0}^{(i)}\hat{\rho}\hat{\sigma}_{0\downarrow}^{(i)} - \frac{1}{2} \left\{\hat{n}_0^{(i)},\hat{\rho}\right\} \right)
\end{split}
\end{align}
with a spontaneous decay rate $\gamma$.
The dynamics of the $N$-body density matrix $\hat{\rho}$ of the lattice is determined by the quantum master equation
\begin{equation}\label{eq:MasterEquation}
\frac{\partial}{\partial t}\hat{\rho}= \text{i} [\hat{\rho},\hat{H}]+\hat{\mathcal{L}}(\hat{\rho}).
\end{equation}
%
%
Its exact solution can, however, be found numerically only for small particle numbers, such that suitable approximations are required to analyze the nonequilibrium physics of large lattices. 

The simplest approach is to entirely neglect correlations by performing a mean-field approximation, which corresponds to the assumption that many-body observables factorize
\begin{align}\label{eq:MF}
\langle\hat{\sigma}_{\alpha\beta}^{(i)}\hat{\sigma}_{\bar{\alpha}\bar{\beta}}^{(j)}\rangle\approx \sigma_{\alpha\beta}^{(i)}\ \sigma_{\bar{\alpha}\bar{\beta}}^{(j)}.
\end{align}
This approach furnishes a set of coupled mean-field equations for the one-body expectation values $\sigma_{\alpha\beta}^{(i)}=\langle\hat{\sigma}_{\alpha\beta}^{(i)}\rangle$ that can be readily solved and have been used to study steady-state phases in different open quantum systems \cite{Lee2011,marcuzzi2014,chan2015limit,owen2018quantum,carollo2022exact,kongkhambut2022observation,wu2024dissipative,li2024time,solanki2024exotic,nadolny2025nonreciprocal,yang2025emergent}. In many cases, the nonlinearity that arises from the mean-field interaction can give rise to bistabilities \cite{Lee2011,carr2013nonequilibrium,de2016intrinsic,vsibalic2016driven} that can often be related to a first-order nonequilibrium phase transition of the exact quantum dynamics \cite{weimer2015}. As pointed out in \cite{wu2024dissipative}, interacting multi-level systems can also support the formation of nonstationary phases, where the system settles into a state with persistent periodic oscillations in the long-time limit. Figure~\ref{fig:pd}a shows the parameter region where we find such a CTC phase with broken time translational symmetry from a linear stability analysis of the mean-field equations in the thermodynamic limit (where we can exploit translational invariance) obtained from Eqs.(\ref{eq:MasterEquation}) and (\ref{eq:MF}) \cite{strogatz2018nonlinear,suppl}. Indeed, the basic idea of CTC phases in an open quantum many-body system \cite{iemini2018boundary} is often based on the validity of mean-field theory. In the present case, this would require $\kappa\le 2$ such that interactions are long-ranged in two dimensions and the mean-field contribution to the interaction dominates in the thermodynamic limit \cite{mattes2025long}. 

Consequently, mean-field theory does not straightforwardly apply to the physics of neutral-atom arrays, where relevant interactions, such as van der Waals ($\kappa=6$) and dipole-dipole ($\kappa=3$) interactions, are shorter-ranged. 
Various methods have recently been explored \cite{weimer21} to study the dissipative dynamics of quantum many-body systems beyond simplified mean-field models.
Here, we employ two complementary approaches to account for correlations in lattices with dipole-dipole interactions ($\kappa=3$). 

First, we use a cumulant expansion that permits a systematic inclusion of correlations through a hierarchy of equations for multi-particle cumulants of increasing order \cite{kubo1962generalized,lackner2017high,Plankensteiner2022,donsa2023nonequilibrium,Verstraelen23,hammerer23}. We take $2^{\rm nd}$-order correlations into account
\begin{align}\label{eq:cumulant}
\langle\hat{\sigma}_{\alpha\beta}^{(i)}\hat{\sigma}_{\bar{\alpha}\bar{\beta}}^{(j)}\rangle= \sigma_{\alpha\beta}^{(i)}\sigma_{\bar{\alpha}\bar{\beta}}^{(j)}+\langle\hat{\sigma}_{\alpha\beta}^{(i)}\hat{\sigma}_{\bar{\alpha}\bar{\beta}}^{(j)}\rangle_c.
\end{align}
but neglect three-particle cumulants to truncate the hierarchy of equations \cite{suppl}. Second, we perform cluster mean-field simulations \cite{jin2016cluster}, whereby the density matrix is factorized  
\begin{align}
\hat{\rho}=\bigotimes_i \hat{\rho}_{\mathcal{C}_i}
\end{align}
into finite clusters, $\mathcal{C}_i$, of particles. Using this ansatz in the master equation (\ref{eq:MasterEquation}), one obtains the exact quantum dynamics on each plaquette $\mathcal{C}_i$, while interactions between particles in different clusters are included on a mean-field level. In the thermodynamic limit, one can assume homogeneity such that all plaquettes become identical, $\hat{\rho}_{\mathcal{C}_i}=\hat{\rho}_{\mathcal{C}}$ \cite{suppl}. We have performed simulation for $2\times2$ as well as $3\times3$ clusters and find very similar results for the parameters of Figures ~\ref{fig:pd}(b)-(e) (see \cite{suppl}).

While either method is approximate, they provide complementary approaches to include quantum fluctuations. In the cumulant expansion, we account only for direct binary correlations but do so for all particle distances throughout the lattice. On the other hand, the cluster mean-field simulations treat correlations only within a finite interaction range but include all orders of correlations via an exact solution of the master equation within a chosen plaquette $\mathcal{C}$. Through this complementarity, the application and comparison of both methods affords more stringent conclusions about the emergence of quantum time-crystalline phases.

In order to detect oscillatory solutions in such simulations we use the time-averaged variance  %
\begin{align}
\label{eq:osc}
\delta_\uparrow^2=\frac{1}{\tau}
\int_t^{t+\tau}\!\!\!\!n_\uparrow(t^\prime)^2{\rm d}t^\prime-\left(\frac{1}{\tau}
\int_t^{t+\tau}\!\!\!\!n_\uparrow(t^\prime){\rm d}t^\prime\right)^2
\end{align}
of the Rydberg-state density, $n_\uparrow=N^{-1}\sum_i\langle\hat{n}^{(i)}_\uparrow\rangle$, in the long-time limit ($t\rightarrow\infty$) and increase $\tau$ until $\delta_\uparrow^2$ is converged. By construction, $\delta_\uparrow^2$ vanishes for stationary steady states and yields a measure of the oscillation amplitude of $n_\uparrow$ in the time-crystal phase.

\begin{figure}[b!]
\includegraphics[width=\linewidth]{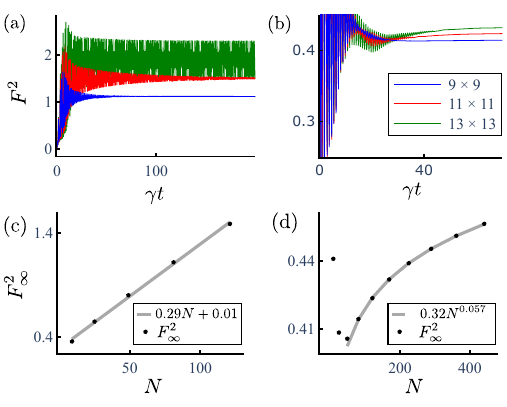}
\caption{\label{fig:quantum_flucts} Dynamics of the normalized variance $F^2=\Delta N_\uparrow^2/N$ in the (a) qCTC-I and the (b) qCTC-II phase. Panels (c) and (d) show the corresponding dependence of steady-state value $F^2_\infty=\lim_{t \rightarrow \infty} F^2$ on the particle number $N$. The interaction strength was fixed to $\chi = -18 \gamma$ for all system sizes. Fits to a linear (c) and a power law (d) are shown in gray. Since the longer-lived oscillations in qCTC-I prevent reaching a steady state within reasonable simulation times for large systems, we show less data points in (c).}
\end{figure}

Figure~\ref{fig:pd}(a) shows how the phase diagram is modified upon including correlations via the $2^{\rm nd}$-order cumulant expansion. For most parameters, the boundary time-crystal phase predicted by mean-field theory is expectedly damped upon inclusion of correlations, as illustrated in Figure~\ref{fig:pd}(e). Remarkably, however, we find a small but finite parameter region where the time-crystalline phase persists in the beyond-mean-field dynamics. We denote this form of quantum time crystals as qCTC-I and have confirmed their existence both via the cumulant expansion and cluster mean-field simulations, as shown in Figure~\ref{fig:pd}(d) \cite{suppl}. Surprisingly, we find another time-crystal phase, marked as qCTC-II in Figure~\ref{fig:pd}(a). This type-II oscillatory solution emerges outside the classical mean-field regime, i.e., in a parameter region where the mean-field solution settles into a stationary steady state \cite{suppl}. Rather than damping the mean-field dynamics, accounting for correlations causes a breaking of time-translation symmetry. Oscillations in this novel time-crystal phase are thus driven by beyond-mean-field effects, as shown in Figure~\ref{fig:pd}(b).

For a more detailed analysis of the two qCTC phases, we study the variance 
\begin{align}
    \Delta N_\uparrow^2 = \langle\hat{N}_\uparrow^2\rangle - \langle\hat{N}_\uparrow\rangle^2,
\end{align}
of the total population in the $\lvert\uparrow\rangle$-state, given by $\hat{N}_\uparrow = \sum_i \hat{n}_\uparrow^{(i)}$. For boundary time crystals with all-to-all interactions \cite{iemini2018boundary}, pertinent to mean-field theory, it was found that fluctuations of the type $F^2=\Delta N_\uparrow^2/N$ \cite{benatti2018quantum, goderis1989central, goderis1989non,goderis1990dynamics, verbeure2010many, benatti2016non, benatti2017quantum} grow linearly with the number $N$ of particles while remaining finite for stationary steady states \cite{carollo2022exact}. As shown in Figures~\ref{fig:quantum_flucts}(a) and (c), this behavior is recovered in the qCTC-I phase of our system. In stark contrast, however, in the qCTC-II phase, we find a sublinear growth of $F^2$ upon increasing the particle number $N$. While the observed increase of $F^2$ still distinguishes the time crystal from the stationary phase, the observed weak power-law dependence $F^2\sim N^{0.06}$ differs distinctively from the linear mean-field scaling of the qCTC-I time crystal \cite{suppl}. Notably, this implies that the relative uncertainty $\Delta N_\uparrow/N_\uparrow$ approaches a finite value as $N\rightarrow\infty$ for the type-I time crystal while it vanishes in the thermodynamic limit of the qCTC-II phase. This behavior is consistent with the results of \cite{medenjak2020isolated}, where it was shown that persistent oscillations can emerge in quenched isolated quantum many-body systems in the absence of long-range spatial correlations, demonstrating that spatial long-range order is irrelevant for time-translational symmetry breaking.
\begin{figure}[t!]
\includegraphics[width=\linewidth]{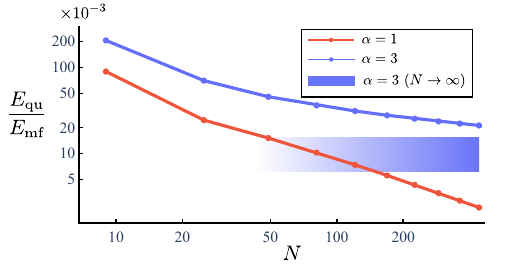}
\caption{\label{fig:energies} Ratio between correlation and mean-field energy for a long ($\kappa = 1$) and a short-range ($\kappa = 3$) interaction potential. Parameters are taken from qCTC-II. In the long-range case, the energy ratio tends to 0 as the thermodynamic limit is approached. In the short-range case, on the other hand, the correlations remain non-negligible as $N \rightarrow \infty$, and the energy ratio fluctuates between finite values due to the broken time-translation symmetry. The oscillation range in the thermodynamic limit for $\kappa = 3$ is depicted as a blue area. For all system sizes and potentials, the interaction strength was fixed to $\chi = -18 \gamma$.}
\end{figure}

Despite the sublinear scaling behavior in qCTC-II, quantum fluctuations have a significant effect on the dynamics of the system in either phase, owing to the finite-range nature of the interaction. We can quantify the impact of correlations by comparing the total mean-field interaction energy
\begin{align}\label{eq:Emf}
E_{\rm mf}=\sum_{i,j>i} U_{ij} n_\uparrow^{(i)}n_\uparrow^{(j)}
\end{align}
to the correlation energy 
\begin{align}\label{eq:Equ}
E_{\rm qu}=\sum_{i,j>i} U_{ij} \langle \hat{n}_\uparrow^{(i)}\hat{n}_\uparrow^{(j)}\rangle_c.
\end{align}
As shown in Figure~\ref{fig:energies} for the qCTC-II phase, the ratio $E_{\rm qu}/E_{\rm mf}$ indeed approaches a finite value in the thermodynamic limit \cite{suppl}.
In contrast, an exact description via mean-field theory requires $E_{\rm qu}/E_{\rm mf}$ to vanish in the thermodynamic limit, which is the case for long-range interacting systems \cite{owen2018quantum,zhu2019dicke,carollo2022exact,kongkhambut2022observation,li2024time,nadolny2025nonreciprocal}. This is illustrated numerically in Figure~\ref{fig:energies} for a long-range interacting lattice with $\kappa=1$ and otherwise identical parameters. As correlations in open quantum systems can be both of classical and quantum origin, we studied the dynamics of the logarithmic negativity, which provides a measure of entanglement \cite{vidal2002computable}. Indeed, we find finite entanglement in both time-crystal phases \cite{suppl}. Thus, we conclude that the qCTC-II time-crystal phase can be considered a quantum effect beyond the classical dynamics of dissipative time crystals under mean-field conditions. 

In summary, we have studied the dissipative dynamics of interacting three-level quantum systems and found nonequilibrium phases with persistent oscillations that extend the mean-field phenomenology of CTC phases. In particular, we have shown that continuous time-translational symmetry can be broken as a direct consequence of beyond-mean-field effects. Fluctuations in such quantum continuous time crystals (here termed qCTC-II) exhibit distinctly different scaling behavior than found in mean-field models with long-range interactions \cite{carollo2022exact}. 

The reported dynamics of spin-$1$ lattices could not be found in equivalent spin-$1/2$ systems, raising interesting questions regarding the fundamental requirements for the emergence of CTC phases and motivating further explorations of minimal models for qCTCs in many-body systems. Addressing these issues could expand the understanding of the basic mechanisms for CTC formation in open quantum systems, possibly beyond the phenomenology of periodic orbits in classical nonlinear systems, and may yield a suitable classification of distinct time-crystalline phases \cite{sieberer2023universality,daviet2024kardar}. Moreover, the observation of persistent oscillations in the presence of local decay poses exciting questions concerning the general stability of different time-crystal phases with respect to perturbations and the role of dissipation \cite{yousefjani2024non}.

Our model applies directly to the physics of neutral-atom arrays \cite{Labuhn2016b,bernien2017probing,scholl2021,browaeys2020many,kaufman2021quantum,srakaew2023subwavelength,weckesser2024realization}, which should make detailed experimental explorations of the predicted time-crystalline dynamics possible. Here, the qCTC-II phase appears particularly attractive as the intermediate-state population turns out to be very small in the oscillating phase. The type-II time crystal, thus, forms an approximate dark state in which persistent oscillations primarily occur between the $\lvert\downarrow\rangle$- and $\lvert\uparrow\rangle$-state, such that heating of the atoms due to photon emission from the $\lvert 0\rangle$-state remains at a low level \cite{suppl}. The parameters considered in this work are typical for arrays of rubidium atoms. The considered Rabi frequencies of around $\sim10\gamma$ correspond to readily achievable values of $\sim13$MHz for two-photon excitation of $nS$-Rydberg states via the intermediate $6P$-state \cite{levine2018high}. 
The chosen interaction strength of $\chi=-18\gamma$ would correspond to a modest nearest-neighbor interaction of $C_3/a^3=-2.6$MHz, which can be realized experimentally by off-resonant microwave dressing \cite{sevinccli2014microwave} of a $\lvert nS\rangle$-Rydberg state with a $\lvert nP\rangle$-state, enabling a strong resonant dipole-dipole interaction. \\

We thank Fan Yang, Klaus Mølmer, Jan Kumlin, Simon P. Pedersen, and Andreas Nunnenkamp for insightful discussions. This work was supported by funding from the Austrian Science Fund (Grant No. 10.55776/COE1) and the European Union (NextGenerationEU), by the SNSF through the Swiss Quantum Initiative, and from the European Research Council through the ERC Synergy Grant SuperWave (Grant No. 101071882).\\

\emph{Note added:} During preparation of this manuscript, we became aware of a related work on time-crystal phases in open spin chains with power-law interactions \cite{wang2025boundary}.

\bibliography{bibliography}
\end{document}


\preprint{APS/123-QED}

\title{\Large Supplementary Material \\[12pt] \textbf{\large Quantum Dissipative Continuous Time Crystals}}

\author{Felix Russo\orcidlink{0009-0005-8975-2245}}
 \email{felix.russo@tuwien.ac.at}
\author{Thomas Pohl}%
 \email{thomas.pohl@itp.tuwien.ac.at}
\affiliation{%
Institute for Theoretical Physics, Vienna University of Technology (TU Wien), Vienna, Austria
}%
\maketitle
\date{\today}

\section{Cumulant Expansion}
\label{app:cumulant}
We work in the Heisenberg picture, i.e., we perform the time evolution of the atomic operators consisting of $\sigma^{(i)}_{\alpha \beta}(t)$ and $\hat{n}^{(i)}_{\alpha}(t) = \sigma^{(i)}_{\alpha \alpha}(t)$, where $\sigma^{(i)}_{\alpha \beta}(0) = \lvert\alpha\rangle_i\langle\beta\rvert$ acts on the $i$th atom and $\alpha, \beta \in \{ \downarrow, 0, \uparrow \}$. The master equation that governs the dynamics is
\begin{eqnarray}
\label{eq:general_eom}
    \text{i} \partial_{t}  \langle \hat{A} \rangle = \langle [\hat{A},\hat{H}] \rangle + \text{i} \langle \mathcal{L}^*(\hat{A}) \rangle,
\end{eqnarray}
for an arbitrary operator $\hat{A}$, where 
\begin{align}
    \mathcal{L}^*(\hat{A}) = \gamma \sum_{i} \left(\hat{\sigma}^{(i)}_{0 \downarrow} \hat{A} \hat{\sigma}^{(i)}_{\downarrow 0} - \frac{1}{2} \{\hat{n}_0^{(i)},\hat{A}\} \right)
\end{align}
is the dual of the Lindbladian from Eq.(2). Using a general notation, one can write the equation of motion for operator products as
\begin{align}
\label{eq:multisite-eom}
\begin{split}
    &\text{i} \partial_{t} \langle \hat{\sigma}^{(i_1)}_{\alpha_1 \beta_1} \hat{\sigma}^{(i_2)}_{\alpha_2 \beta_2} ... \hat{\sigma}^{(i_N)}_{\alpha_N \beta_N} \rangle \\
    &= \Bigg( \langle \hat{\sigma}^{(i_2)}_{\alpha_2 \beta_2} ... \hat{\sigma}^{(i_N)}_{\alpha_N \beta_N} \left( \hat{K}^{(i_1)}_{\alpha_1+1 \beta_1 \alpha_1 \alpha_1 \beta_1 \beta_1} - \hat{K}^{(i_1)}_{\alpha_1 \beta_1-1 \beta_1-1 \alpha_1 \beta_1 \alpha_1} \right) \rangle 
    \\ & \hphantom{=}
    + \sum_{j \neq 1} (i_j, \alpha_j, \beta_j) \leftrightarrow (i_1, \alpha_1, \beta_1) \Bigg)
    \\ & \hphantom{=}
    - \text{c.c.} (\alpha_1 \leftrightarrow \beta_1, \alpha_2 \leftrightarrow \beta_2, ..., \alpha_N \leftrightarrow \beta_N)
\end{split}
\end{align}
where
\begin{align}
\label{eq:operatorK}
\begin{split}
    \hat{K}^{(i)}_{\alpha \beta \mu \nu \rho \tau} 
    &\equiv  \delta_{\beta \mu}  \left( \Delta + \text{i} \delta_{\mu \downarrow} \frac{\gamma}{2} \right) \hat{\sigma}^{(i)}_{\alpha \mu+1} - \Omega_{\mu} \hat{\sigma}^{(i)}_{\alpha \beta}
    \\ & \hphantom{=}
    + \frac{1}{2} \delta_{\tau \uparrow} \sum_{j \neq i} U_{ij} \hat{\sigma}^{(i)}_{\nu \rho} \hat{n}^{(j)}_{\uparrow}.
\end{split}
\end{align}
and we introduced addition of the spin indices $0 = \ \downarrow + 1 = \ \uparrow - 1$. Operators containing indices outside of the ladder are defined to vanish, e.g., $\hat{\sigma}^{(i)}_{\alpha (\uparrow+1)} = \hat{0} \ \forall \alpha, i$.
Moreover, we introduced the helper quantity 
\begin{align}
    \Omega_{\mu} = \begin{cases}
        \Omega \text{ if } \mu = \downarrow \\
        \omega \text{ if } \mu = 0 \\
        0 \text{ else.}
    \end{cases}
\end{align}
From Eq.(\ref{eq:multisite-eom}), one can readily derive an equation of motion for the $2^{\rm nd}$-order cumulants:
\begin{align}
\begin{split}
    &\text{i} \partial_{t} \langle\hat{\sigma}^{(i)}_{\alpha \beta} \hat{\sigma}^{(i')}_{\alpha' \beta'}\rangle_\text{c} 
    \\
    &=\Bigg( \langle\hat{\sigma}^{(i')}_{\alpha' \beta'} \hat{K}^{(i)}_{\alpha+1 \beta \alpha \alpha \beta \beta}\rangle - \sigma^{(i')}_{\alpha' \beta'} \langle\hat{K}^{(i)}_{\alpha+1 \beta \alpha \alpha \beta \beta}\rangle 
    \\ & \hphantom{=} \hphantom{(}
    - \left( \langle\hat{\sigma}^{(i')}_{\alpha' \beta'} \hat{K}^{(i)}_{\alpha \beta-1 \beta-1 \alpha \beta \alpha}\rangle 
    - \sigma^{(i')}_{\alpha' \beta'} \langle\hat{K}^{(i)}_{\alpha \beta-1 \beta-1 \alpha \beta \alpha} \rangle \right) 
    \\ & \hphantom{=} \hphantom{(}
    + (\alpha \leftrightarrow \alpha', \beta \leftrightarrow \beta', i \leftrightarrow i') \Bigg) 
    \\ & \hphantom{=}
    - \text{c.c.} (\alpha \leftrightarrow \beta, \alpha' \leftrightarrow \beta').
\end{split}
\end{align}
Since $\hat{K}$ contains $2^{\rm nd}$-order terms, the $2^{\rm nd}$-order cumulants depend on $3^{\rm rd}$-order expectation values. To obtain a closed set of equations, we approximate the latter by two-body terms, neglecting the third-order cumulant: 
\begin{align}
\begin{split}
    &\langle\hat{\sigma}^{(n_1)}{} \hat{\sigma}^{(n_2)}{} \hat{\sigma}^{(n_3)}{}\rangle
    \\ &
    \approx  
    \langle\hat{\sigma}^{(n_1)}{} \hat{\sigma}^{(n_2)}{}\rangle \langle\hat{\sigma}^{(n_3)}{}\rangle  
    + \langle\hat{\sigma}^{(n_1)}{} \hat{\sigma}^{(n_3)}{}\rangle  \langle\hat{\sigma}^{(n_2)}{}\rangle  
    \\ &\hphantom{=}
    + \langle\hat{\sigma}^{(n_1)}{}\rangle  \langle\hat{\sigma}^{(n_2)}{} \hat{\sigma}^{(n_3)}{}\rangle  
    -2 \langle\hat{\sigma}^{(n_1)}{}\rangle  \langle\hat{\sigma}^{(n_2)}{}\rangle  \langle\hat{\sigma}^{(n_3)}{}\rangle ,
\end{split}
\end{align}
where we omitted the spin indices.
\subsection{Thermodynamic Limit}
\label{app:cumulant-td-limit}
\begin{figure}
    \includegraphics[width=\linewidth]{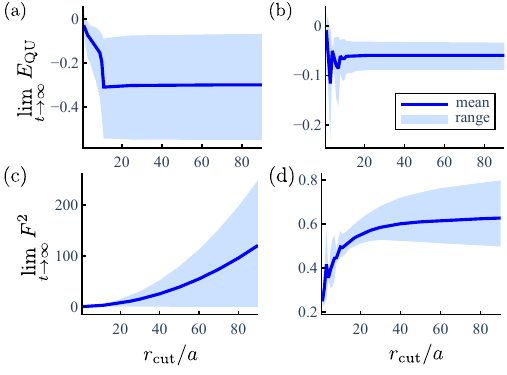}
    \caption{\label{fig:cutoff_analysis} (a), (b) Oscillation range of the (a)/(b) the correlation energy and (c)/(d) the normalized fluctuations in the long-time limit as a function of the cut-off radius in the thermodynamic limit. The panels on the left, i.e.,(a) and (c), show parameters within qCTC I while the panels on the right, i.e.,(b) and (d), show parameters within qCTC II.  While the correlation energies remain finite, the fluctuations diverge. The kink in (a) comes from a more attractive periodic orbit that the observables start to follow.}
\end{figure}   
In the thermodynamic limit, we can make use of the homogeneity of the system to only simulate one set of single-site equations corresponding to an arbitrary atom, which we label with the index $0$ in the following. However, the single-site operators are coupled to infinitely many two-site operators, see Eq.(\ref{eq:operatorK}) for example. Performing a cumulant expansion (cf. Eq.(5)), the interaction term of the single-site operators can be rewritten
\begin{align}
\label{eq:interaction-td-limit}
     \sum_{j \neq 0} U_{0 j}{} \langle\hat{\sigma}^{(0)}_{\alpha \beta} \hat{n}_\uparrow^{(j)}\rangle 
    &= \chi \sigma^{}_{\alpha \beta} n_{\uparrow}{} + \sum_{j \neq 0} U_{0 j}{} \langle\hat{\sigma}^{(0)}_{\alpha \beta} \hat{n}_\uparrow^{(j)}\rangle_\text{c},
\end{align} 
where we used the translational symmetry in the thermodynamic limit and introduced $\sigma^{}_{\alpha \beta} \equiv \sigma^{(0)}_{\alpha \beta}$. From Eq.(\ref{eq:interaction-td-limit}), one sees immediately that the mean-field equations are nonlinear and agnostic to the microscopic details of the potential (they only depend on the effective nonlinearity $\chi$). 

To simulate the dynamics in the thermodynamic limit despite the infinite number of $2^{\rm nd}$-order correlations appearing in Eq.(\ref{eq:interaction-td-limit}), we use that the Rydberg potential decays as a function of the distance between sites, allowing us to cut off the sum appearing in the second term. In Figure~\ref{fig:cutoff_analysis}(a) and (b), we show the oscillation ranges of the correlation energy (i.e., the second term in Eq.(\ref{eq:interaction-td-limit}) for $\alpha = \beta = \ \uparrow$, see Eq.(10)) in the limit of long simulation times as a function of where we choose the cutoff. For both time-crystalline phases, we find that (1) the correlation energy oscillates between finite values and (2) the sum has converged when cutting it off after $r_{\text{cut}} = 40 a$, where $a$ is the lattice constant. 

As discussed in the main text, the normalized fluctuations 
\begin{align}
\begin{split}
    F^2 
    &= n_{\uparrow}{} - n_{\uparrow}^{2} + \sum_{j \neq 0}  \langle\hat{n}_\uparrow^{(0)} \hat{n}_\uparrow^{(j)}\rangle_\text{c}, \\
\end{split}
\end{align}
diverge as a function of $N$ for finite systems and, therefore, also do not converge as function of $r_{\rm cut}$, as seen in Figure~\ref{fig:cutoff_analysis}(c) and (d).
\section{Cluster Mean-Field Approach}
\label{app:cluster-mf}
\begin{figure}
    \includegraphics[width=\linewidth]{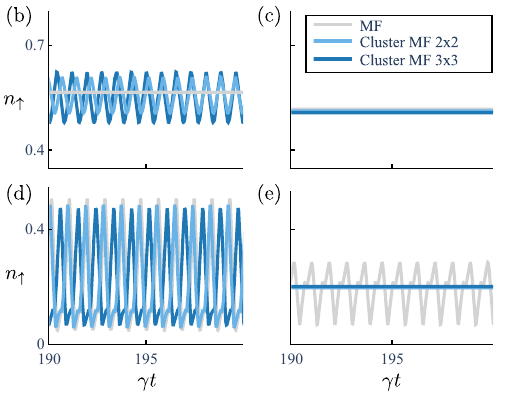}
    \caption{\label{fig:clustermf} Dynamics of the spin-up population at large simulation times for different cluster sizes. The parameters in (b)-(d) correspond to the four points labeled with these letter in the phase diagram in Figure~2a. The cluster mean-field simulations are in qualitative agreement, deviating from the simple mean-field prediction.}
\end{figure}   
Another approach to include short-range correlations is the cluster mean-field approach. 
Hereby, the lattice is grouped into small connected clusters $\mathcal{C}_i$ that are assumed to factorize, i.e.,
\begin{align}
    \hat{\rho} = \otimes_i \hat{\rho}_{\mathcal{C}_i}.
\end{align}
Substituting this relation into the master equation and tracing over all clusters except one (denoted by $\mathcal{C}$), we obtain 
\begin{align}
\label{eq:cluster-master-eq}
\begin{split}
    \text{i} \partial_{t} \hat{\rho}_\mathcal{C} &= \sum_{j \in \mathcal{C}} \left([\hat{h}_{j},\hat{\rho}_\mathcal{C}] + \text{i} \hat{\mathcal{L}}_{j}(\hat{\rho}_\mathcal{C}) \right)
    \\ &\hphantom{=}
    + \frac{1}{2} \sum_{j, k \in \mathcal{C}} U_{jk}{} [\hat{n}_\uparrow^{(j)} \hat{n}_\uparrow^{(k)},\hat{\rho}_\mathcal{C}]
    \\ &\hphantom{=}
    + \sum_{j \in \mathcal{C}, k \notin \mathcal{C}} U_{jk}{} [\hat{n}_\uparrow^{(j)} n_{\uparrow}^{(k)},\hat{\rho}_\mathcal{C}]
\end{split}
\end{align}
where we used the cyclic property of the trace and introduced the single-particle Hamiltonian 
\begin{align}
\begin{split}
    \hat{h}_{i} &= - \Delta \left( \hat{n}_0^{(i)} + \hat{n}_\uparrow^{(i)}\right) + \left(\Omega \hat{\sigma}^{(i)}_{\downarrow 0} + \omega \hat{\sigma}^{(i)}_{0 \uparrow}  + \text{h.c.} \right)
\end{split}
\end{align}
as well as the single-particle Lindbladian
\begin{align}
\begin{split}
    \hat{\mathcal{L}}_{(i)}(\hat{A}) = \gamma \left(\hat{\sigma}^{(i)}_{\downarrow 0} \hat{A} \hat{\sigma}^{(i)}_{0 \downarrow} - \frac{1}{2} \{\hat{n}_0^{(i)},\hat{A}\} \right).
\end{split}
\end{align}
While the first two lines in Eq.(\ref{eq:cluster-master-eq}) give the exact solution for the cluster $\mathcal{C}$, the term in the last line accounts for the mean-field interaction with the atoms in the remaining clusters. Practically, this means that we simulate the exact dynamics of the cluster including a self-consistent detuning
\begin{align}
\begin{split}
    \tilde{\Delta}_{j} = \sum_{k \notin \mathcal{C}} U_{jk}{} n_{\uparrow}^{(k)},
\end{split}
\end{align}
where $j \in \mathcal{C}{}$. It is intuitively clear that we approach the exact solution as the cluster size grows. Due to the three-level structure of our particles, we are limited to clusters of size $2 \times 2$ and $3 \times 3$. However, the dynamics corresponding to the two cluster sizes are in qualitative agreement, see Figure~\ref{fig:clustermf}.
\section{Comparison of the Spectra}
\begin{figure}
    \includegraphics[width=\linewidth]{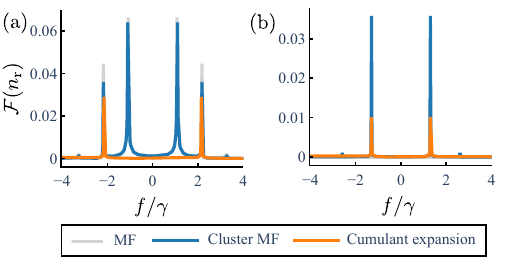}
    \caption{\label{fig:spectra} Fourier spectra of the spin-$\uparrow$ population dynamics for parameters from qCTC I (a) (cf. Figure~2(d) in the main text) and qCTC II (b) (cf. Figure~2(b)).}
\end{figure}  
Despite the vastly different approximations that are performed in the cumulant expansion and the cluster mean-field approach, the spectra of the corresponding dynamics show similar features. In Figure~\ref{fig:spectra}, we compare the Fourier components of the mean-field and the two beyond mean-field approaches for parameters from both time-crystalline phases. We find that, for both phases, the beyond mean-field techniques share a dominant frequency component.
\section{Trap Heating by Spontaneous Decay}
\begin{figure}
    \includegraphics[width=\linewidth]{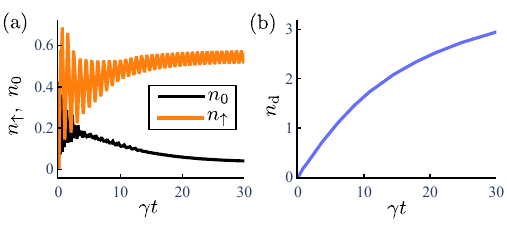}
    \caption{\label{fig:heating} (a) Dynamics of the Rydberg and the intermediate-state population at large simulation times as predicted by a cumulant expansion. The intermediate-state population oscillates with an amplitude that is not resolved in this plot. (b) Number of decays per atom as a function of time. Parameters are taken from the qCTC II phase.}
\end{figure}   
One of the main sources of loss in a tweezer experiment comes from heating due to decays of the intermediate state. A measure for the amount of dissipation is the number of decays per atom, i.e., in the thermodynamic limit,
\begin{align}
    n_\text{d}(t) = \gamma \int_0^t n_{0}(\tau) d\tau.
\end{align}
As shown in Figure~\ref{fig:heating}(a), the population of the intermediate state is quite low in the qCTC II phase. Consequently, few decays happen and one should be able to observe multiple oscillations of the Rydberg population before the losses due to heating lead to a deterioration of the experimental data, see Figure~\ref{fig:heating}(b). Note that the population of the intermediate state also oscillates in Figure~\ref{fig:heating}(a) but with a smaller amplitude.
\section{Negativity in the Time-Crystalline Phases}
Negativity is a measure for entanglement that can readily be computed via the partial transpose of a density matrix $\hat{\rho}^{T_i}$ with respect to the $i^{\rm th}$ spin. If the partially transposed density matrix has negative eigenvalues, correlations in the system are necessarily quantum, i.e., the system features entanglement. 
For a given many-body density operator $\hat{\rho}$, the negativity can be computed as 
\begin{align*}
    \mathcal{N}(\hat{\rho}) = \frac{\rvert \rvert \hat{\rho}^{T_i}\lvert\lvert_1 - 1}{2},
\end{align*}
\begin{figure}
    \includegraphics[width=\linewidth]{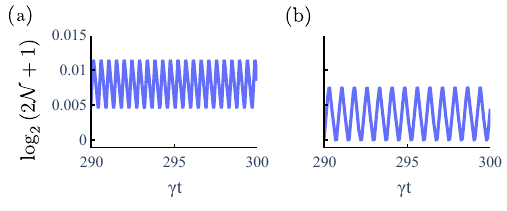}
    \caption{\label{fig:neg} Logarithmic negativity of the partially transposed two-body reduced density matrix for neighboring spins. The calculations are performed in the thermodynamic limit using the second-order cumulant expansion. We consider parameters within qCTC I (a) and qCTC II (b), corresponding to Figures 2(b) and 2(d) of the main text.}
\end{figure}  
where the trace norm is defined as $\rvert \rvert  \hat{\rho} \lvert \lvert_1 = \text{Tr}\left(\sqrt{\hat{\rho}^\dagger \hat{\rho}}\right)$. Essentially, $\mathcal{N}$ corresponds to the sum of all negative eigenvalues of the density matrix, partially transposed with respect to the $i^{\rm th}$ spin. To explore the existence of entanglement in the qCTC phases we consider two-body reduced density matrices, which we obtain directly from the second-order cumulant approach. Taking the partial transpose with respect to one spin of a given pair and calculating the corresponding negativity gives direct access to the two-body entanglement in the system. In Figure~\ref{fig:neg}, we show the 
logarithmic negativity 
\begin{align*}
    {\rm log}_2\rvert \rvert \hat{\rho}_{ij}^{T_i}\lvert\lvert_1={\rm log}_2\left(2\mathcal{N}(\hat{\rho}_{ij})+1\right)
\end{align*}
of a the two-body reduced density matrix $\hat{\rho}_{ij}$ for two nearest neighbors, ${\bf r}_i$ and ${\bf r}_j$, in the lattice.
The calculations are performed for the same parameters as in Figures 2(b) and 2(d) of the main text, and demonstrate that the two time crystalline phases, qCTC-I and qCTC-II, indeed feature finite entanglement.

\section{Linear Stability Analysis}
Having identified a fixed point $\mathbf{x_0}$ in a system of nonlinear equations, 
\begin{align}
    \frac{\partial}{\partial t} \mathbf{x} = \mathbf{f}(\mathbf{x})
\end{align}
one can determine its stability by investigating the eigenvalues of the system's Jacobian evaluated at the fixed point 
\begin{align}
    \mathbf{J} = \frac{\partial \mathbf{f}}{\partial \mathbf{x}} \Bigg \rvert_{\mathbf{x_0}}.
\end{align}
If the Jacobian possesses an eigenvalue with a positive real part, the fixed point is unstable, i.e., a small perturbation would push the system dynamics away from $\mathbf{x_0}$. Moreover, one can use the eigenvalues of $\mathbf{J}$ to determine whether a supercritical Hopf bifurcation has occurred, i.e., whether a stable fixed point became repulsive while an attractive limit cycle emerged. This happens when a pair of complex conjugate eigenvalues crosses the imaginary axis and thus obtains positive real parts.

In the theory of nonlinear systems, a Hopf bifurcation is not the only generation mechanism of attractive limit cycles. Therefore, a linear stability analysis is in general not sufficient to identify all stable periodic orbits. For the mean-field equations in the thermodynamic limit (where we can exploit the translational invariance) and the considered parameters, however, the only limit cycles stem from Hopf bifurcations as we checked by performing time evolutions from random initial conditions at different points in the phase diagram. For more details on nonlinear systems refer to.
\section{Absence of a mean-field limit cycle in the qCTC-II regime}
\begin{figure}[h!]
    \includegraphics[width=\linewidth]{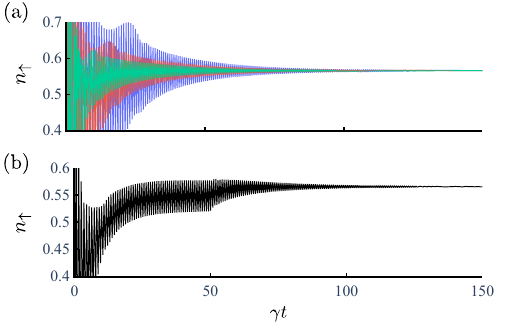}
    \caption{\label{fig:absence_mf_lc} Dynamics of the $\lvert \uparrow \rangle$-state population in the thermodynamic limit for qCTC-II parameters. (a) Mean-field evolution from three random initial conditions. A stationary steady state is reached independently of the initial condition. (b) Time evolution using a second-order cumulant expansion. At $\gamma t = 50$, all two-particle cumulants are set to zero and the system is evolved using a mean-field approach. This leads to the damping of the time-crystalline dynamics.}
\end{figure}

To ensure that there is no mean-field limit cycle that is not found by the linear stability analysis, we perform mean-field simulations from random initial conditions using parameters from the qCTC-II regime. We find that a stationary steady state is reached independently of the initial condition, see Figure~\ref{fig:absence_mf_lc}(a). Another way to check that there is no mean-field limit cycle for qCTC-II parameters is to perform a second-order cumulant expansion and, once the time-crystalline phase is reached, turn off quantum effects by dropping all second-order cumulants (i.e., by evolving the system further using a mean-field approach). As shown in Figure~\ref{fig:absence_mf_lc}(b), the qCTC-II phase does not survive in the absence of beyond-mean-field terms.
\newpage
\bibliography{bibliography}